\documentclass{egpubl}
\usepackage{eurovis2024}

\EuroVisShort  %

\CGFccby

\usepackage[T1]{fontenc}
\usepackage{dfadobe}  

\usepackage{cite}  %
\BibtexOrBiblatex
\electronicVersion
\PrintedOrElectronic
\ifpdf \usepackage[pdftex]{graphicx} \pdfcompresslevel=9
\else \usepackage[dvips]{graphicx} \fi

\usepackage{egweblnk}
\usepackage{colortbl}
\usepackage{multirow}
\usepackage{graphicx}
\usepackage{subfigure}
\usepackage{makecell}
\usepackage{appendix}

\makeatletter
\let\orgdescriptionlabel\descriptionlabel
\renewcommand*{\descriptionlabel}[1]{%
  \let\orglabel\label
  \let\label\@gobble
  \phantomsection
  \edef\@currentlabel{#1\unskip}%
  \let\label\orglabel
  \orgdescriptionlabel{#1}%
}
 \makeatother
 
\title[Revisiting Categorical Color Perception in Scatterplots]%
      {Revisiting Categorical Color Perception in Scatterplots: \\ Sequential, Diverging, and Categorical Palettes\vspace{-1.5em}}

\author[C. Tseng et al.]
{\parbox{\textwidth}{\centering Chin Tseng$^{1}$, Arran Zeyu Wang$^{1}$, Ghulam Jilani Quadri$^{1,2}$, and Danielle Albers Szafir$^{1}$ 
        }
        \\
{\parbox{\textwidth}{\centering $^1$University of North Carolina at Chapel Hill, NC, USA\\
         $^2$University of Oklahoma, OK, USA
       }
}
}

\begin{document}

\teaser{
  \vspace{-4em}
  \includegraphics[width=0.85\linewidth]{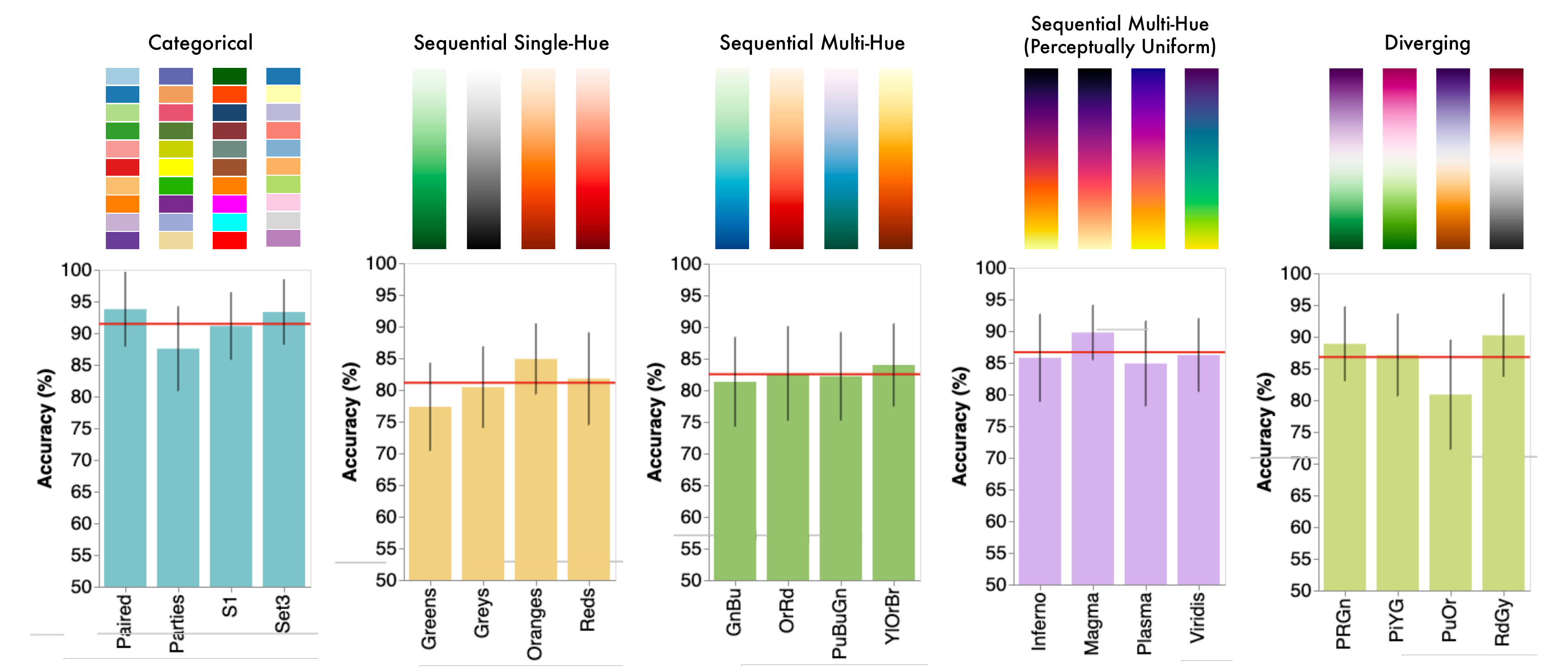}
  \vspace{-1em}
  \centering
  \caption{The accuracy rates for relative mean judgment tasks separated per color palette design family, with their average accuracy per family (red line). 
  We find that design families significantly influenced class comparison, with multi-hue categorical palettes providing the highest performance (91.44\%), followed by diverging (86.78\%),  perceptually-uniform sequential palettes (86.67\%), multi-hue sequential palettes (82.56\%), and single-hue sequential palettes (81.11\%).
  }
  \label{fig:teaser}
}

\maketitle

\begin{abstract}
Existing guidelines for categorical color selection are heuristic, often grounded in intuition rather than empirical studies of readers' abilities.
While design conventions recommend palettes maximize hue differences, more recent exploratory findings indicate other factors, such as lightness, may play a role in effective categorical palette design. We conducted a crowdsourced experiment on mean value judgments in multi-class scatterplots using five color palette families---single-hue sequential, multi-hue sequential, perceptually-uniform multi-hue sequential, diverging, and multi-hue categorical---that differ in how they manipulate hue and lightness. Participants estimated relative mean positions in scatterplots containing 2 to 10 categories using 20 colormaps. Our results confirm heuristic guidance that hue-based categorical palettes are most effective. However, they also provide additional evidence that scalable categorical encoding relies on more than hue variance.


\vspace{-0.5em}
\end{abstract}

\section{Introduction}
\label{sec-intro}
\vspace{-0.5em}
People use scatterplots for a range of tasks~\cite{szafir2016four,quadri2021survey,quadri2021constructing}, including comparing trends~\cite{harrison2014ranking}, exploring clusters~\cite{quadri2020modeling,quadri2022automatic,jeon2023clams}, and estimating values~\cite{gleicher2013perception, hong2021weighted}.
Designers commonly use mark color to encode categorical data in scatterplots~\cite{gleicher2013perception}.
However, 
categories become harder to compare as 
the number of categories increases~\cite{tseng2023evaluating}. 
Intentional design choices can provide robust data interpretation as the number of categories increases; however, most palette design guidelines are heuristic, often grounded in designer intuition rather than quantitative evidence. 

One common guideline for categorical palette design is to maximize hue variation \cite{munzner2014visualization, gramazio2016colorgorical, reda2018graphical, stone2006choosing}.
Tools like ColorBrewer \cite{harrower2003colorbrewer} or Tableau offer categorical encodings that embody this idea, first emphasizing nameable color differences, then varying lightness and saturation as secondary factors to extend palettes to larger numbers of colors. 
The principles 
of effectiveness and expressiveness \cite{munzner2014visualization, mackinlay1986automating} further suggest that palettes emphasize hue variation over lightness to avoid unintentionally implying order within a palette and to align with human intuition. However, recent studies indicate that these principles may not apply universally or uniformly 
\cite{tseng2023evaluating}. 
Instead, they contend lightness variation can strengthen categorical perception, but draw these conclusions by studying only palettes with strong baseline hue variation. 
This study aims to examine traditional design heuristics for categorical color encoding empirically. 

We investigate these heuristics in 
multiclass scatterplots, measuring 
performance across five design families of palettes characterized by their respective use of hue and lightness: categorical, single-hue sequential, multi-hue sequential, perceptually-uniform multi-hue sequential, and diverging (\autoref{fig:teaser}). 
Classical visualization guidelines encourage the use of hue-varying categorical palettes and discourage other families that rely more heavily on lightness. However, recent exploratory analyses found that hue variation alone may not be the strongest predictor of robustness in categorical color encoding design \cite{tseng2023evaluating}.
Our approach explores this tension in a class averaging task previously used to investigate the role of color.
We employed the task of judging the highest mean y-values in multiclass scatterplots \cite{gleicher2013perception,tseng2023evaluating}.

We conducted a crowdsourced experiment to explore the impact of sequential, diverging, and categorical color palettes on categorical perception.
Our experimental results reveal that both enlarging hue variance (two-hue diverging or multi-hue sequential) and emphasizing perceptual uniformity can benefit 
categorical perception and lead to more robust performance over larger numbers of categories. 
We reconcile our results with current empirical design guidelines grounded in studies of exclusively categorical palettes \cite{tseng2023evaluating}, concluding that while hue variation dominates performance, lightness variation and general perceptual distance also play a notable role in categorical encoding design.

\vspace{-1em}
\section{Related Work}
\label{sec-related}
We briefly review related literature about categorical perception in scatterplots and color palette design and perception.

\vspace{-1em}
\subsection{Categorical Perception in Scatterplots}

Categorical perception~\cite{goldstone2010categorical} describes how %
people perceive different groups (or categories) of objects. 
Existing studies assessed categorical perception in scatterplots for various tasks, such as measuring the influence of shape, size, and color~\cite{szafir2018modeling, smart2019measuring, burlinson2017open}, and approximating difference metrics between categorical channels
\cite{demiralp2014learning}. 
However, the emphasis on minimum differentiable thresholds in these studies characterizes pairwise interactions between marks; the results may not generalize to holistic relationships across an entire palette.

Subsequent studies have explored people's abilities to compare values between larger numbers of categories. An initial experiment focused on comparing mean values in scatterplots, indicating that performance remained consistent as the number of categories increased from two to three~\cite{gleicher2013perception}.
More recently, Tseng et al.~\cite{tseng2023evaluating} further extended the mean estimation task to more complex scatterplots with up to ten categories, finding that both the number
of categories and discriminability of colors significantly impact 
people's abilities to compare categories.
However, these studies focused on a limited number of colors with an emphasis on color palettes that primarily varied in hue. 
Tseng et al.'s exploratory analysis suggested that lightness may play a larger role than hue in predicting palette robustness across varying category numbers. We extend this observation to consider a wider range of color design choices that vary along both hue and lightness.

\vspace{-1em}
\subsection{Color Palette Design and Perception}

Prior work offers a range of design guidelines and measures for effective colormap design \cite{bujack2017good}.
For example, we can compute the \emph{perceptual distance} between two colors 
using metrics from color science like CIEDE2000~\cite{sharma2005ciede2000} or from visualization~\cite{szafir2018modeling}. 
Guidelines further note that hue variation and lightness difference specifically impact people's color palette interpretation~\cite{tseng2023evaluating, kindlmann2002face}.
As the difference between colors increases, perceptual distance metrics become less useful. However, measures such as color name difference \cite{heer2012color} provide a metric for color uniqueness or relatedness for larger color differences \cite{gramazio2016colorgorical}. 
Color palettes also must consider people's aesthetic preferences, 
which, contrary to categorical color design recommendations, may increase with hue similarity~\cite{schloss2011aesthetic, palmer2013visual}.
For example, \emph{pair preference} models human preference using aesthetic color appearance factors including lightness and saturation~\cite{schloss2011aesthetic}.

These metrics inform broader design guidelines for assembling sets of colors into colormaps for representing data. 
Perceptually-uniform colormaps improved accuracy and response time when comparing values~\cite{liu2018somewhere}. 
Diverging colormaps support more robust pattern perception than sequential colormaps in heatmaps~\cite{reda2018graphical}.
Reda \& Szafir~\cite{reda2020rainbows} 
found that increased color name variability improves graphical inference. 
Color design tools like 
Colorgorical \cite{gramazio2016colorgorical} leverage these findings using a set of color perception, aesthetic, and naming models to score palettes.

Effective color use goes beyond task alignment.
For example, people more effectively interpret visualizations when colors align with their corresponding semantic concepts~\cite{schloss2018color,kinateder2019color}. 
Schloss et al.~\cite{schloss2018mapping} 
found that people use implicit color associations to infer values from colormaps, such as dark-is-more or opaque-is-more biases. 
Follow-up studies suggest semantics are as important as perceptual measures in categorical visualization~\cite{mukherjee2021context}.

Despite numerous perceptual measures 
relevant to categorical color palette design, 
most guidelines for constructing color ramps are loosely-defined qualitative heuristics, posing challenges for users without professional experience~\cite{smart2019color}. 
We take a step towards better bridging heuristic and perceptual measures in categorical colormap design by investigating hue, lightness, and uniformity variations across different palette design families.

\vspace{-1em}
\section{Methodology}
\label{sec-method}

We analyzed how 
the color palettes used to distinguish 
categories impact people's abilities to reason with multiclass scatterplots. We conducted a crowdsourced study measuring how well people 
compare category means over varying category numbers ($N=2-10$) and five color palette families (\autoref{fig:teaser}). These families systematically vary in hue and lightness, representing common perceptual properties used in palette design. 
Given the importance of overall perceptual discriminability 
and uniformity between categories, we additionally tested perceptually-uniform sequential palettes to understand whether uniformity may play a role in categorical encoding, where color differences between marks tend to be larger than in standard unit steps with continuous encodings. 
We hypothesized:

\noindent\textbf{H1:} \textbf{Categorical palettes will perform better than all other palettes.}
Categorical palettes are designed with best practices and principles for categorical data encoding. These palettes emphasize hue variations while conforming to effectiveness and expressiveness principles for categorical data~\cite{munzner2014visualization}, unlike diverging or sequential palettes designed to communicate quantitative data. 

\noindent\textbf{H2:} \textbf{Diverging palettes will perform better than sequential palettes.}
Diverging palettes encourage people to cluster data by using hue to indicate the end and middle points of a distribution. This grouping may reduce people's abilities to estimate quantitative statistics within a palette compared to sequential encodings~\cite{liu2018somewhere}; however, this reduction may indicate stronger grouping cues than sequential palettes with similar hue variation which outperform diverging palettes on quantitative estimation tasks.

\noindent\textbf{H3:} \textbf{Multi-hue sequential palettes will perform better than single-hue sequential palettes.}
Prior experiments reported using multi-hue palettes to represent data could 
help people distinguish data and find correlations~\cite{liu2018somewhere, spence1999using, levkowitz1996perceptual, borland2007rainbow}.
Given hue's importance in palette design heuristics, we anticipate palettes varying more in hue will lead to better performance. 

\noindent\textbf{H4:} \textbf{Perceptually-uniform sequential palettes will perform better than other sequential palettes.}
Many popular sequential palettes are perceptually non-uniform \cite{smart2019color}. 
Perceptual distance metrics help align 
colors with human perception~\cite{gomez2016comparison, tseng2023evaluating} and have commonly been applied in color measurement and palette design~\cite{gramazio2016colorgorical, heer2012color, sharma2005ciede2000}.
We anticipate that more uniform steps between colors will improve people's abilities to distinguish colors across palettes by ensuring a sufficient minimum difference between pairs of categories and 
perceived category differences that are proportional to the number of categories.

\vspace{-1em}
\subsection{Task \& Stimuli}
\begin{figure}[b] 
\vspace{-1em}
\centering
\includegraphics[width=0.45\textwidth]{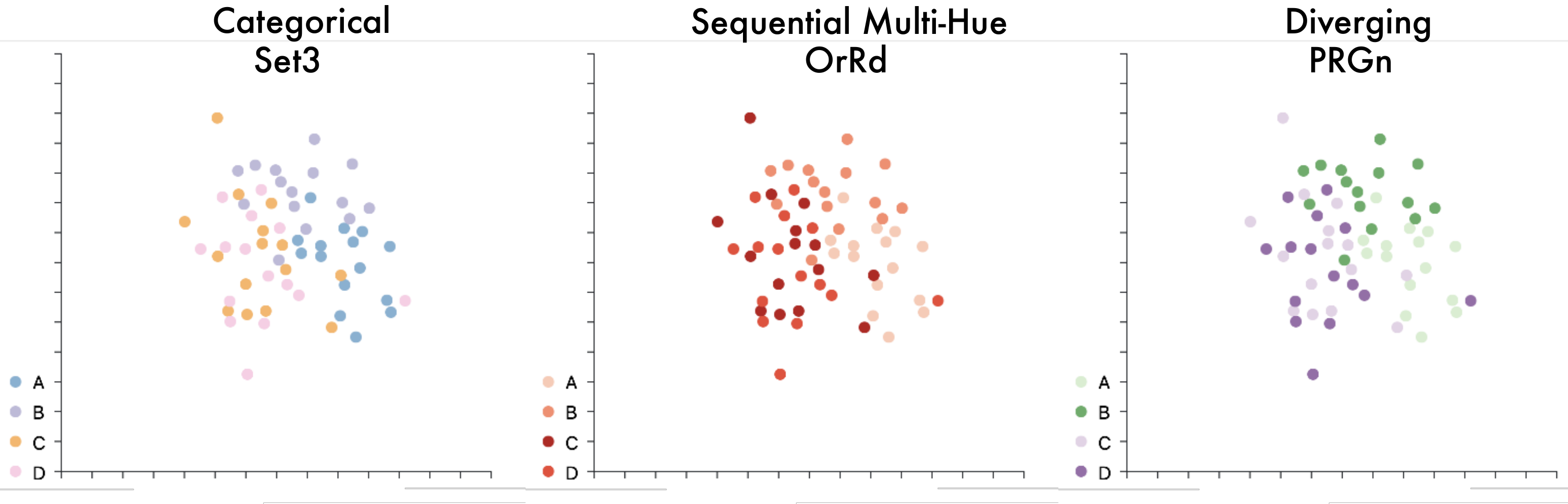} 
\vspace{-0.5em}
\caption[]{Example stimuli for three palette types.}
\vspace{-1em}
\label{fig:stimuli}
\end{figure}
We used a relative mean judgment task from previous studies~\cite{gleicher2013perception,szafir2016four,hong2021weighted,tseng2023evaluating}. As in Tseng et al.~\cite{tseng2023evaluating}, we asked participants to estimate the category with the highest average y-value (e.g., ``\emph{Which category has the highest average mean value?}''). 
This task required participants to first find the data points of each category and then estimate statistical values over all points in that category such that confusion between categories (colors) leads to errors in estimation. 

We generated each scatterplot as a $400{px}\times 400{px}$ graph using D3.
Scatterplots were rendered on a white background using two orthogonal black axes with 13 unlabeled ticks. 
We generated point positions according to the data distribution methods in Tseng et al. ~\cite{tseng2023evaluating} with a ``medium'' hardness level and 15 points with a four-pixel radius-filled circle in each category to generate 180 datasets.
We selected 20 color palettes for five color families (four per family). \autoref{fig:teaser} shows the palettes and \autoref{fig:stimuli} the example stimuli.
We chose these color palettes from expert-crafted examples 
based on
their performance in previous studies~\cite{smart2019color,tseng2023evaluating}, 
their high relative lightness magnitude or variance~\cite{zeileis2009escaping}, and high \textit{perceptual distance} between colors~\cite{sharma2005ciede2000}.
For categorical palettes, colors were randomly selected. We sampled colors from continuous palettes by mapping them to the range [0, 1] and using uniform data steps to choose the corresponding categorical colors.

\vspace{-1em}
\subsection{Procedure \& Participants}
Our study involved three phases: (1) informed consent and colorblindness screening, (2) task description and tutorial, and (3) formal study. Participants initially gave consent per IRB guidelines, provided demographics, completed an Ishihara test for colorblindness, and then were introduced to the tasks. 
Participants clicked on a data point in the category with the highest average y-value in a scatterplot. We randomly divided the 20 color palettes into 4 groups with each group containing one palette from each family. Each participant responded to 45 stimuli containing one combination of each palette from certain group $\times$ category number (2--10) presented in random order with a random dataset. We randomly assigned the group when participants started the study.
To avoid learning or fatigue effects, three engagement checks were randomly placed along with the formal questions. The engagement checks had three categories with large differences in their means. 

We recruited 112 U.S. and Canada-based participants via MTurk. Data from 100 participants (63 male, 37 female, aged 24-65) without CVD and passing all engagement checks were analyzed, with each receiving \$3.00 compensation. The anonymized data, stimuli, results, and infrastructure 
for our study can be found on \href{https://osf.io/scqjb/?view_only=77c68f9aa476446f973451db7792c0e0}{OSF}.

\vspace{-1em}
\section{Results}
\label{sec-results}
\begin{figure*}[htbp] 
\vspace{-0.5em}
\centering
\includegraphics[width=0.95\textwidth]{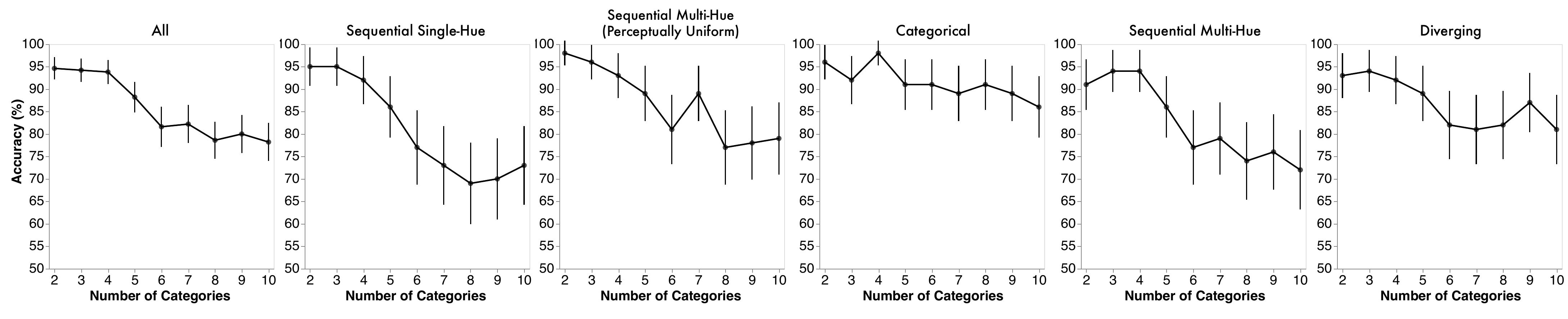} 
\vspace{-1em}
\caption[]{Accuracy rates broken down by category number separated by color palette families. We found that categorical palettes were more robust to increasing numbers of categories, with other families dropping in performance for six or more categories. }
\vspace{-1.5em}
\label{fig:category-acc}
\end{figure*}

Accuracy was the primary dependent measure. To compare performance across palette designs and category numbers, we conducted a two-factor ANOVA with color palette family and number of categories as independent variables and Tukey’s HSD with $\alpha = 0.05$ and Bonferroni correction for post-hoc analysis. 

\vspace{3pt}
\noindent\textbf{Color Families: }
Our results show that color palette families 
significantly impact performance
($F(4, 95)=12.25,
p<.0001$).
As shown in \autoref{fig:teaser}, 
categorical palettes  (91.44\% accuracy) significantly outperformed diverging (86.78\%), perceptually-uniform sequential multi-hue (86.67\%), sequential multi-hue (82.56\%), and sequential single-hue palettes (81.11\%), supporting \textbf{H1}. 
Diverging and perceptually-uniform multi-hue palettes
outperformed sequential single-hue palettes.
These findings partially support \textbf{H2}: diverging palettes outperformed single-hue sequential palettes. 

Our results partially support \textbf{H3}: perceptually-uniform multi-hue sequential palettes 
performed better than single-hue sequential palettes;
however, perceptually-uniform multi-hue palettes did not outperform
general multi-hue sequential palettes. 
In partial support of \textbf{H4}, perceptually uniform multi-hue sequential palettes performed marginally better than non-uniform multi-hue palettes. However, further analysis revealed that perceptually uniform palettes traversed a larger range of color differences than non-uniform multi-hue palettes. Given the larger range of differences and predictable difference structure imparted by uniformity, one might expect perceptually-uniform palettes to dramatically outperform non-uniform palettes, contrary to the relatively small difference in our findings. Part of this difference may pertain to degradation in perceptual distance metrics at larger differences \cite{heer2012color,gramazio2016colorgorical}. Future work should explore the role of perceptual uniformity and related metrics in categorical palette design.

\vspace{3pt}
\noindent\textbf{Number of Categories:}
Our results confirm Tseng et al.'s findings~\cite{tseng2023evaluating} that \textit{performance decreased as the number of categories increased}.
Our results reveal a significant effect of the number of categories on judgment accuracy ($F(8, 91)=20.68, p<.0001$): 
people were less accurate with higher numbers of categories (\autoref{fig:category-acc}).
For every family, the accuracy rate decreased as the number of categories increased.
However, the categorical palettes remained relatively robust 
even with higher numbers of categories. 
Sequential and diverging palettes remained notably robust to small numbers of categories 
(less than six).

\vspace{-1em}
\section{Discussion}
\label{sec-discussion}
When designing categorical encodings, our results empirically confirm heuristic guidance that hue is the most significant factor. However, other factors also likely play a role in encoding robustness. 
Key findings and design implications from our study include:

\vspace{3pt}
\noindent \textbf{Categorical palettes 
effectively encode categorical data.}
Predominantly hue-varying categorical color palettes achieved the highest and most robust 
accuracy among all palette types. 
This finding validates heuristic best practices for categorical encoding design.
Further, it provides evidence in support of the expressiveness and effectiveness principles \cite{munzner2009nested, quadri2024do}: palettes that maximize intuitive categorical attributes outperformed those that maximize ordinal attributes (i.e., sequential and diverging palettes).  

\vspace{3pt}
\noindent \textbf{Hue is the primary factor for effective palette design, but may be insufficient when considered alone. }
The average performance and robustness of multi-hue sequential palettes were higher than single-hue. This finding aligns with several existing studies~\cite{liu2018somewhere, spence1999using, levkowitz1996perceptual, borland2007rainbow}, which indicate multi-hue sequential palettes make it easier to find correlations between certain colors. 

However, our results also indicate that hue alone is insufficient to explain performance. For example, mean overall performance and robustness for the single-hue and non-uniform multi-hue color palettes were nearly identical despite multi-hue palettes having greater hue variation, more nameable hues, and larger overall perceptual distance between colors. 
Diverging palettes performed slightly better overall and were more robust than sequential palettes with as many or more distinct hues. These results, coupled with findings from previous exploratory studies that indicate the importance of lightness variation in categorical encodings~\cite{tseng2023evaluating}, suggest an interplay between lightness, hue, and categorical perception. These variations may also relate to perceived ordering or similar factors. Understanding these factors and their interactions is important future work.

\vspace{3pt}
\noindent \textbf{Increasing category numbers influences performance.}
Our experiments support that adding more categories 
significantly reduces categorical perception accuracy~\cite{tseng2023evaluating}. Further, these findings partially replicate the subitizing phenomenon found in past work by revealing a noticeable performance drop between five and six classes~\cite{tseng2023evaluating, haroz2012capacity}.
In our case, performance appears to fall rapidly between five and six categories but remains relatively stable for six to ten categories. This stability may indicate a change in the perceptual strategy used to process larger numbers of categories that should be interrogated in future work. 

Our results provide mixed evidence of how palette design might influence this threshold. We did not find a significant drop in categorical palettes; however, diverging palettes, which are essentially two single-hue sequential palettes joined by white in our study, exhibited the same ``S''-shaped dip in performance at six categories as sequential palettes, with slightly higher average performance after six categories. Exploring this tension and how design might mitigate (or leverage) changing perceptual strategies in categorical data visualization is important future work. 

\noindent \textbf{Limitations \& Future Work:}
This study examines design factors in categorical encodings using a small set of expert encodings for mean estimation.
While largely consistent with Tseng et al.'s findings~\cite{tseng2023evaluating}, discrepancies in categorical palette performance rankings, possibly due to varied study settings, require further analysis.
Contrary to expectations, perceptually-uniform palettes with larger average color step sizes did not significantly outperform non-uniform ones, suggesting a need for more research into the impact of perceptual metrics on palette design.
Future work should also 
explore the potential misinterpretation of category ordering in sequential encodings %
to more holistically characterize color palette design's role in expressiveness and effectiveness.

\vspace{-1em}

\section{Acknowledgments}
This work was supported by the National Science Foundation under grant No. 2127309 to the Computing Research Association for the CIFellows project, NSF IIS-2046725, and NSF IIS-1764089.

\clearpage

\bibliographystyle{eg-alpha-doi}
\bibliography{main}

\end{document}